\documentclass[11pt]{article}

\usepackage{array}
\usepackage{amsmath,amsbsy}
\usepackage{amsxtra}
\usepackage{amstext}
\usepackage{amssymb}
\usepackage{pdfpages}
\usepackage{times}
\usepackage{hyperref}
\usepackage{fullpage}
\usepackage{algorithmic,booktabs,bm,rotating}
\usepackage{url}

\def\E{\mathop{\rm E\,\!}\nolimits}

\def\tr{\mathop{\rm tr}\nolimits}

\def\vec{\mathop{\rm vec}\nolimits}

\newcommand{\by}{\boldsymbol{y}}

\newcommand{\bA}{\boldsymbol{A}}

\newcommand{\bD}{\boldsymbol{D}}

\newcommand{\bI}{\boldsymbol{I}}

\newcommand{\bO}{\boldsymbol{O}}

\newcommand{\bR}{\boldsymbol{R}}
\newcommand{\bS}{\boldsymbol{S}}

\newcommand{\bU}{\boldsymbol{U}}
\newcommand{\bV}{\boldsymbol{V}}
\newcommand{\bW}{\boldsymbol{W}}
\newcommand{\bX}{\boldsymbol{X}}
\newcommand{\bY}{\boldsymbol{Y}}
\newcommand{\bZ}{\boldsymbol{Z}}

\newcommand{\bbeta}{\boldsymbol{\beta}}

\newcommand{\bsigma}{\boldsymbol{\sigma}}

\newcommand{\bSigma}{\boldsymbol{\Sigma}}

\newcommand{\bPhi}{\boldsymbol{\Phi}}

\newcommand{\mendel}{{\sc Mendel}}

\newcommand{\openmendel}{{\sc OpenMendel}}

\newcommand{\julia}{{\sc Julia}}

\renewcommand{\vec}[1]{\mathbf{#1}}

\newcommand{\Ibf}{{\bm I}}

\newcommand{\Ybf}{{\bm Y}}

\newcommand{\zerobf}{{\mathbf 0}}

\begin{document}

\title{\openmendel: A Cooperative Programming Project for Statistical Genetics} 

\author{
    Hua Zhou$^{\text{\textbf{1,2}}}$ \thanks{Department of Biostatistics, UCLA Fielding School of Public Health, Los Angeles, CA, USA} \and
    Janet S. Sinsheimer$^{\text{\textbf{1,2}}}$ \thanks{Department of Human Genetics, David Geffen School of Medicine at UCLA, Los Angeles, CA, USA} \and
    Douglas M. Bates \thanks{Department of Statistics, University of Wisconsin, Madison, Wisconsin, USA} \and
    Benjamin B. Chu \thanks{Department of Biomathematics, David Geffen School of Medicine at UCLA, Los Angeles, CA, USA} \and
    Christopher A. German\footnotemark[1] \and
    Sarah S. Ji\footnotemark[1] \and 
    Kevin L. Keys \thanks{Department of Medicine, University of California, San Francisco, CA, USA} \and
    Juhyun Kim\footnotemark[1] \and 
    Seyoon Ko \thanks{Department of Statistics, Seoul National University, Seoul, South Korea} \and
    Gordon D. Mosher \thanks{Departments of Statistics and Computer Science, University of California, Riverside, CA, USA} \and
    Jeanette C. Papp \footnotemark[2] \and
    Eric M. Sobel \footnotemark[2] \and 
    Jing Zhai \thanks{Department of Epidemiology and Biostatistics, Mel and Enid Zuckerman College of Public Health, University of Arizona, Tucson, AZ, USA} \and
    Jin J. Zhou \footnotemark[8] \and 
    Kenneth Lange$^{\text{\textbf{2}}}$ \footnotemark[4] \\
    $^{\text{\sf 1}}$ These authors contributed equally. \\
    $^{\text{\sf 2}}$ To whom correspondence should be addressed. \\
    \textbf{Email}: \texttt{huazhou@ucla.edu}, \texttt{jsinshei@g.ucla.edu}, \texttt{klange@ucla.edu}  \\
    \textbf{Phone}: +1 (310) 794-7835, +1 (310) 825-8002, +1 (310) 206-8076
    \date{\today}
}

\maketitle

\begin{abstract}
Statistical methods for genomewide association studies (GWAS) continue to improve.
However, the increasing volume and variety of genetic and genomic data make computational speed and ease of data manipulation mandatory in future software.
In our view, a collaborative effort of statistical geneticists is required to develop open source software targeted to genetic epidemiology.
Our attempt to meet this need is called the \openmendel project (\url{https://openmendel.github.io}).
It aims to (1) enable interactive and reproducible analyses with informative intermediate results, (2) scale to big data analytics, (3) embrace parallel and distributed computing, (4) adapt to rapid hardware evolution, (5) allow cloud computing, (6) allow integration of varied genetic data types, and (7) foster easy communication between clinicians, geneticists, statisticians, and computer scientists.
This article reviews and makes recommendations to the genetic epidemiology community in the context of the \openmendel\ project. \\

    \textbf{Keywords}: Statistical Genomics, GWAS, Computational Statistics, Open Source, Collaborative Programming
\end{abstract}

\section{Introduction}

Genomewide association studies (GWAS) query the entire genome to identify genetic variants associated with a trait of interest. GWAS have enjoyed many successes \cite{visscher201710} and have uncovered many clues to the genetic etiology of common diseases \cite{cookson2009mapping}.  Case-control tests of association between markers and traits predate GWAS by more than 50 years \cite{aird1953relationship}.  However, association studies were rarely undertaken in the pre-GWAS era unless there were candidate genes with strong prior evidence. The situation changed at the turn of the millennium when dense SNP (single nucleotide polymorphism) maps became available and SNP genotyping costs plummeted. Suddenly, it became possible to exploit linkage disequilibrium (LD) and survey hundreds of thousands to millions of genomewide SNPs. In the subsequent decade, hundreds of associations found by GWAS were published \cite{visscher2012five,visscher201710}. Genomics is in the midst of a second technological evolution driven by high-throughput sequencing \cite{metzker2010sequencing,pickrell2012next,kilpinen2013next,van2014ten}.  Geneticists can now survey both rare and common variants.

This sudden expansion of data leads to enormous challenges in statistical genetics. Many current algorithms and programs are ill adapted to handle modern data sets with $10^5$ cases and $10^7$ markers. Ever more types of genetic variation are being observed and catalogued \cite{visscher201710}. These changes demand more complex data structures and data integration across multiple biological scales. Precision health and predictive medicine raise the stakes even further \cite{kilpinen2013next}. Concurrently, the nature of computing is rapidly changing. In addition to new hardware, new programming paradigms and new algorithms must be brought online as quickly as possible to sustain progress in statistical genetics. 

The following three studies exemplify the variety and magnitude of genomic data sets being collected today: a) The  Million Veteran Program contains GWAS data (657,459 SNPs) on 359,964 veterans \cite{Gaziano2016MVP}. Simply storing the genotypes in compressed format requires $>100$ GB. Obviously, this data set and others like it \cite{sudlow2015uk,brody2017analysis} will continue to grow. b) A recent study \cite{Telenti2016HumanLongevity} obtained whole genome sequence (WGS) data on 10,545 humans at 30-40x  coverage for $<$ \$2000 per genome. These researchers identified $>$ 150 million variants, the majority of which are rare or \textit{de novo}. c) The iPOP (integrative personal omics profile) study \cite{Chen2012iPOP} followed a single individual for 401 days and collected transcriptome, proteome, metabolome, microbiome, epigenome, exposome, and phenome data at 20 time points, along with an extremely high coverage WGS. This type of omics profiling yields a dynamic picture of the heteroallelic changes between healthy and diseased states. 

Current analysis pipelines juggle a multitude of computer programs that are implemented in different languages, run on different platforms, and require  different input/output formats. This heterogeneity unintentionally creates barriers to communication, data exchange, data visualization, and scientific replication. End-users treat the entire pipeline as a black box and often fail to use their biological insight to inform statistical analysis. Students, post-docs, and researchers spend inordinate amounts of time coding and debugging the low-level languages instead of thinking about the science. In addition to these disadvantages, current software packages are straining under the volume, velocity, variety, and veracity of modern genomics data. Many programs do not even run on multiple threads. Distributed computing across different machines is largely ignored. In our view, the time is ripe to put in place a better paradigm for statistical genetics.

In this review, we first explain why the new \julia\ language \cite{BezansonEdelmanKarpinskiShah17Julia} is an ideal choice for the \openmendel\ analysis platform. We then present what we see as some of the most pressing needs in gene mapping and our efforts to advance them through the cooperative \openmendel\ effort. Owing to page and time constraints, we do not offer encyclopedic coverage of recent advances in GWAS or sequence analysis.  Many promising methods are left unmentioned, for example meta-analysis based on summary statistics \cite{cantor2010prioritizing,Chiu16FDAPleiotropy,Fan16FDAMeta,kim2015adaptive,mancuso2017integrating,yang2012conditional} or estimation of  fine-scale population structure \cite{novembre2016recent}.   Instead we focus on topics related to projects already underway in \openmendel. These projects include methods for handing SNP data, genotype imputation, rapid GWAS, iterative hard thresholding, kinship comparison, variance component modeling, and SNP-set analyses.

We want to point out that there are other groups who have made notable strides in making genetic analyses accessible to researchers who work with big data but lack the support available at large genomic centers  \cite{Bickerstaffe17Ark,Ranaweera18PedDataTools}.  Some of these projects are further along than \openmendel.  A particularly interesting example is the PLATO software project (\cite{hall2017plato} and \url{https://ritchielab.org/software/plato-download}), which is designed to provide a single platform for a variety of association analyses.  A major difference in our approach and PLATO is language choice.  This difference might seem minor but, as we outline below, we believe it is a fundamental difference and is important for our goal of getting user initiated modules and modifications. Another example is the Ark software \cite{Bickerstaffe17Ark}, which focuses on data management and is complementary to rather than competitive with \openmendel. 

\section{The Importance of the Julia Computing Language}
\label{sec:julia}

Many compelling features make \julia\ \cite{BezansonEdelmanKarpinskiShah17Julia} an ideal vehicle for implementing methods for modern statistical genetics. First, it is free, open source, and easy to install. Second, its clear, powerful syntax lends itself to compact, readable code and quick algorithm mock-ups. As needed, it can easily call Fortran, C, R, and Python functions. Third, because \julia\ incorporates an excellent just-in-time (JIT) compiler,  it achieves the efficiency of low-level languages with minimal programming efforts. Fourth, \julia\ is built for parallelism at the multicore, graphical processing unit (GPU), and cluster levels. Fifth, \julia\ employs a modern, easy to use package management system. Of particular relevance to statistical genetics, \julia\ has many statistical and numerical analysis packages ready to use. Finally, end-users can run their analyses via the interactive Jupyter ({\bf Ju}lia, {\bf Py}thon, {\bf R}) Notebook, an attractive interface for data visualization and reproducible research. Together these tools constitute an integrated environment for rapid prototyping of new applications and, with the same code, the analysis of large-scale genetic data. 

Traditional high-level languages such as R, Matlab, and Python face the notorious two-language problem. In this scenario, one high-level language is used for prototyping, but a second low-level language is later needed for producing fast code for real world, large data sets. The high-level code is typically more compact, readable, and amenable to change, but much slower to execute. Most of the popular statistical genetics analysis tools or their most demanding subroutines are implemented purely in low-level languages, greatly restricting the community that feels comfortable exploring the code. Most tools are also restricted to certain computer platforms and input formats. 

Today, a typical analysis pipeline requires a glue language such as Bash, Perl or Python to chain packages together. Data plotting and display require additional software, typically R or Matlab. Current analysis pipelines are cumbersome, opaque, and error-prone, creating barriers to the development of new statistical methods. Researchers wade through a swamp of low-level code and reinvent statistical genetics wheels instead of focusing on their unique contributions. This can be avoided as \julia\ has solved the two-language problem through careful design of the programming language itself. \julia\ is both easy to code and scales to peta-flop computing levels \cite{Claster17Blog}. We can now use \julia\ in all phases of our methods development, from prototyping to production software. \openmendel\ includes many leading-edge statistical genetics methods written in this fast, high-level language that invites easy contributions from scientists. Using \julia, \openmendel\ can become the first highly efficient, open source statistical genetics software that can scale to million-subject studies and is both user- and developer-friendly.

\section{Handling SNP Data}

The SnpArrays.jl module of \openmendel\ provides a convenient bridge between binary SNP data and downstream statistical analysis. The VCFTools.jl module achieves the same end for the richer genetic information distributed in VCF and BCF file formats. In SnpArrays.jl, biallelic genotype data are held in BitArrays, which store four genotypes per byte. As much as possible, compressed storage is also maintained during computation. Julia allows operators such as matrix multiplication to be defined directly on BitArrays without decompression. The design features of Julia make it easy to build high-performance statistical genetics software that is scalable to data sets with millions of subjects and tens of millions of SNPs. 

The functionality of SnpArrays.jl includes: (1) reading and writing compressed SNP files, (2) computing summary statistics, (3) filtering
data by genotyping success rates and other criteria, (4) copying compressed data into numerically oriented vectors and matrices, (5)
computing genetic relationship matrices, (6) computing principal components, and (7) extending matrix and vector operations to compressed SNP data. SnpArrays.jl serves as a data interface to other \openmendel\ modules.

\section{Genotype Imputation}

Genotype imputation involves the inference of unobserved genotypes from observed genotypes. It is possible to base inference on the observed genotypes of surrounding pedigree members \cite{sobel1996haplotyping}, but pedigree data are now viewed as poor substitutes for linkage disequilibrium. In particular, pedigree data are incapable of imputing genotypes at completely untyped SNPs in a study. Recent versions of genotype imputation rely on panels of reference genotypes and employ hidden Markov models, with the hidden states being underlying haplotype pairs \cite{howie2009impute,li2010mach,Marchini2007}.  These programs are computationally intensive and operate by haplotyping individuals on the typed SNPs in the sample. These partial haplotypes are then compared to the reference panel to impute the full set of genotypes \cite{howie2012fast,van2015population}.  We have taken an alternative approach based on the generic data mining technique of matrix completion \cite{CandesRecht09MatrixCompletion,chi2013genotype}. 

Matrix completion fills in the missing entries of an $m \times n$ matrix 
$\bX=(x_{ij})$ whose observed entries are indexed by a
subset $\Omega$ of $\{1, \ldots, m\} \times \{1, \ldots, n\}$. Imputation involves finding a low rank matrix $\bY=(y_{ij})$ consistent with the observed entries of $\bX=(x_{ij})$. This is done by minimizing the loss function 
\begin{eqnarray}
\label{eq:matrix_completion}
f(\bY) & = &  \sum_{(i,j) \in \Omega} (x_{ij} - y_{ij})^2
\end{eqnarray}
over the set of matrices $\bY$ of rank $r$ or less. Taking $r$ small is a form of parsimony capturing the hidden structure of the data. In genotype imputation, $\bX$ records the observed genotype dosages (0, 1, or 2 counts of the reference allele), with rows corresponding to people and columns to  SNPs. Imputation is performed over a narrow genomic window of a few hundred SNPs where linkage disequilibrium 
prevails. Including reference individuals typed on out-of-sample SNPs is
a key part of the strategy.

Because every rank $r$ matrix $\bY$ of dimension $m \times n$ can be expressed as a matrix product $\bU\bV$, where $\bU$ is $m \times r$ and $\bV$ is $r \times n$, matrix completion can be phrased as updating the factors $\bU$ and $\bV$ of $\bY$ in the loss function (\ref{eq:matrix_completion}).
Imputation is iterative, and to restore symmetry at iteration $m$, each missing entry $x_{ij}$ is imputed by its current best guess $(\bU_m\bV_m)_{ij}$. New values $\bU_{m+1}$ and $\bV_{m+1}$ can be recovered by taking the singular value decomposition (SVD) of $\bZ_m$, the current completed version of $\bX$. The MM (majorization/minimization) principle of optimization shows that this procedure drives the loss downhill \cite{Lange16MMBook}.

Chi et al. \cite{chi2013genotype} compared the matrix completion program Mendel Impute to several popular model based imputation programs including MaCH and IMPUTE2 using a number of simulated and real datasets. The accuracy of imputation is dependent on the nature of the specific scenarios and so no program was universally most accurate. The least favorable scenario for Mendel Impute in terms of accuracy occurred when imputing genotypes between high density microarray platforms using as a measure of accuracy the mean $r^2$ between the imputed values and the true genotypes at masked loci. In this case Mendel Impute was slightly worse then MaCH which was slightly worse than IMPUTE2 (Table \ref{table:impute}). In other scenarios Mendel Impute was more accurate than MaCH and IMPUTE2 and in still others they were are roughly the same. However, in all the scenarios presented Mendel Impute was at least an order of magnitude faster than MaCH or IMPUTE2.

\begin{table}
\begin{center}
 \begin{tabular}{c c c c} 
 \toprule
  & Mendel Impute & MaCH & IMPUTE2 \\
 \midrule
 $r^2$ & 0.683 & 0.751 & 0.802 \\ 
 relative time & 1.00 & 13.10 & 7.41 \\
 \bottomrule
\end{tabular}
\end{center}
\caption{Comparison of imputation methods.}
\label{table:impute}
\end{table}

Alternating least squares provides an alternative to SVD that is potentially much faster \cite{HastieMazumderLeeZadeh15ALS}. The alternating updates
\begin{eqnarray*} 
\bV_{m+1} & = & (\bU_m^t\bU_m)^{-1}\bU_m^t\bY_m \quad \quad \text{and}\\
\bU_{m+1} & = & \bZ_m \bV_{m+1}^t (\bV_{m+1}\bV_{m+1}^t)^{-1}
\end{eqnarray*}
can achieve extremely high numerical throughput on modern computer architecture such as multicore CPUs and multiple GPUs. Because alternating
least squares offers no guarantee of finding the global minimum of the loss, initial values for $\bU$ and $\bV$ should be as accurate as possible. Application of a randomized SVD to supply initial values is one possibility \cite{Liberty07rSVD}. In practice we divide the current window into equal thirds and construct a hold-out-set by masking entries in the outer two thirds. We then choose the best rank $r$ based on performance on the hold-out-set. Once we impute missing entries in the middle third, we shift the window to the right and begin again.

\section{Enhancements to Ordinary GWAS}

MendelGWAS.jl performs ordinary SNP-by-SNP association testing. To maximize speed in linear, logistic, and Poisson regression, MendelGWAS.jl employs score tests 
\cite{amin2007genomic,chen2007family,Clark2016MFGScoreTest,Zhou2017FastScore}. For the most significant SNPs, the score test is supplemented by the slower but more accurate likelihood ratio test (LRT). Principal components can be included as predictors, SNPs and subjects can be filtered by success rates, and a Manhattan plot is provided.

In addition to these standard approaches to GWAS, we are in the process of implementing score tests for generalized linear models (GLMs) \cite{schaid2002score}. GLMs permit trait-genotype relations to be modeled  with more exotic response distributions. We are also planning to develop an efficient score test for the challenging Cox survival model  \cite{kawaguchi2018,mittal2013high,suchard2013massive}. Multinomial regression models for complex categorical phenotypes would be a valuable extension of logistic regression \cite{morris2010powerful}. Finally, efficient GWAS for ordered discrete phenotypes is becoming increasingly important for the study of complex diseases and traits derived from electronic health records.

\section{Iterative Hard Thresholding}
\label{Iterative Hard Thresholding}

To avoid the computational complexity of multiple regression and the identifiability issues caused by having more predictors $p$ than sample individuals $n$ \cite{buhlmann2011statistics}, GWAS has traditionally focused on the marginal effects of single SNPs. Previously we introduced lasso penalized regression to GWAS to perform subset selection \cite{wu2009genome,zhou2010association}. Our recent paper \cite{keys2017iterative} implements a better heuristic, iterative hard thresholding (IHT), to solve this inherently combinatorial problem. We showed 
that IHT is better for GWAS than lasso or MCP penalties in controlling for false positive and false negative rates, in reducing parameter shrinkage, and in capturing heritability. It achieves these goals with little sacrifice in computational speed.

We now sketch how IHT iterates toward good local optima. To keep the discussion simple, consider the setting of linear regression with design matrix $\bX$, response vector $\by$, and parameter vector $\bbeta$. The goal is to minimize the loss function $f(\bbeta) = \frac{1}{2}\|\by - \bX\bbeta\|^2$ subject to the sparsity condition $\|\bbeta\|_0 \leq k$. The notation $\|\bbeta\|_0$ is shorthand for the number of nonzero entries of $\bbeta$. In GWAS the entry $x_{ij}$ of $\bX$ denotes the number (0, 1, or 2) of reference alleles carried by individual $i$ at SNP $j$ or the imputed dosage value. The entry $y_i$ of $\by$ corresponding to individual $i$ encodes a continuous trait such as height, blood pressure, or an expression level.

At iteration $n$, the IHT algorithm \cite{blumensath2008iterative} moves in the steepest descent direction $-\nabla f(\bbeta_n)$ modified by the sparsity constraint. Here the gradient $\nabla f(\bbeta)$ of the objective equals $ -\bX^t(\by-\bX\bbeta)$. The IHT update is explicitly
\begin{eqnarray} \label{IHT}
\bbeta_{n+1} & = & P_{S_k}\left(\bbeta_{n} - s \nabla f(\bbeta_n)\right),
\end{eqnarray} 
where $s$ is the steplength and $P_{S_k}(\bbeta)$ denotes projection onto the sparsity set $S_k = \{\bbeta:\|\bbeta\|_0 \le k\}$. The projection operator $P_{S_k}(\bbeta)$ sends to 0 all but the $k$ largest entries of $\bbeta$ in magnitude. The preferred entries of $\bbeta$ are untouched. The steplength $s$ is chosen to minimize $f(\bbeta)$
along the ray $s \mapsto \bbeta_n-s\nabla f(\bbeta_n)$ prior to projection.
This is achieved by taking 
\begin{eqnarray*} \label{IHTstep}
s & = & \frac{\|\nabla f(\bbeta_n)\|^2}{\|\bX \nabla f(\bbeta_n)\|^2}.
\end{eqnarray*}
The best value of $k$ can be chosen by cross-validation.

The theory and practice of IHT continues to advance. Shen and Li \cite{shen2017tight} show how to relax the restricted isometry property originally invoked to prove convergence \cite{blumensath2009iterative}. Yang et al \cite{yang2016selective} suggest group-sparse IHT to promote sparsity on a group-level. Khanna and Kyrillidis \cite{khanna2017iht} validate the application of momentum acceleration to IHT. Yuan et al \cite{yuan2017gradient} and Bahmani \cite{bahmani2013greedy} adapt IHT to logistic regression. Further extension to generalized linear models is a natural target. MendelIHT.jl brings IHT  under the \openmendel\ umbrella. Integration of IHT with SnpArrays.jl unifies data handling and leads to faster code with a smaller memory footprint. Finally, we are investigating weighting predictors to accommodate candidate genes and candidate SNPs \cite{zhou2011penalized}.

\section{Kinship Comparison}

Kinship coefficients quantify the degree of relationship between two relatives. Two genes are identical by descent (IBD) if one is a copy of the other or they are both copies of the same ancestral gene. The theoretical kinship coefficient $\phi_{ij}$ is the probability that a randomly sampled gene at some arbitrary locus from individual $i$ is IBD to a randomly sampled gene at the same locus from individual $j$. For example, if we assume no inbreeding, $\phi_{ij}= \frac{1}{2}$ if $i=j$, and $\phi_{ij}=\frac{1}{4}$ if $i$ and $j$ are first degree relatives. In the former case, the two genes are sampled with replacement. In an accurately constructed pedigree, the full matrix $\bPhi$ of kinship coefficients $\phi_{ij}$ can be calculated from a simple recurrence \cite{lange2003mathematical}. Jacquard's more complex kinship coefficients \cite{jacquard1974genetic} are less useful in practice and harder to calculate \cite{lange1992calculation}. In the MendelKinship.jl module of \openmendel, Jacquard's coefficients are approximated by the Monte Carlo method of gene dropping.

When pedigrees are unknown or suspect, SNP markers can be used to estimate the kinship matrix $\bPhi$ empirically.  One popular estimate is the genetic relationship matrix (GRM), represented here by $\bS=(s_{ij})$. If $p_k$ denotes the reference allele frequency of SNP $k$, $x_{ik}$ counts the number of reference alleles carried by individual $i$, and $K$ is the number of SNPs, then the elements of $\bS$ are calculated as
\begin{eqnarray*}
s_{ij} & = & \frac{1}{K} \sum_{k=1}^K \frac{(x_{ik} - 2p_k)(x_{jk} - 2p_k)}
{4p_k(1 - p_k)}.
\end{eqnarray*}
 Alternatives to the GRM include a methods of moments estimator MoM \cite{day2011linkage} and a robust GRM \cite{manichaikul2010robust,vanraden2008efficient}. The latter is
\begin{eqnarray*}
\hat{\phi}_{ij} & = & \frac{1}{\sum_{k=1}^K 4p_k(1 - p_k)}
\sum_{k=1}^K (x_{ik} - 2p_k)(x_{jk} - 2p_k).
\end{eqnarray*}
This unbiased estimator generally has smaller variance than the standard estimator $\bS$, which is sensitive to low minor allele frequencies \cite{wang2017efficient}.  The MendelKinship.jl module calculates the GRM, the robust GRM, and the MoM estimators. All three of these estimators are special cases of general kinship estimators that are unbiased under ideal conditions \cite{wang2017efficient}. When there is ethnic inhomogeneity and spread in the degrees of relationships, $\bS$ can exhibit bias because it confounds close relatedness and ancestry differences \cite{conomos2016model}.  Ethnic admixture can be accommodated by replacing the allele frequency $p_k$ by an ethnic specific estimate for each individual $i$ \cite{conomos2016model}.
 
Finding the variances of these estimators has been impossible without simplifying assumptions \cite{wang2017efficient}. Our own unpublished approximation to the variance 
\begin{eqnarray}
\E(\|\bS -\E(\bS)\|_{\text F}^2) & \approx & \frac{1}{K^2} \| \bR\|_{\text F}^2 \Big[\|\bPhi\|_{\text F}^2+\tr(\bPhi)^2 \Big] \label{approx_variance}
\label{Var_S}
\end{eqnarray}
of the GRM matrix $\bS$ allows for inbreeding, linkage disequilibrium, and closely related relatives. It relies on the simplifying assumption that the fourth moments of the SNP counts coincide with the fourth moments of similarly distributed Gaussian random variables. In formula (\ref{approx_variance}), $\tr(\bA)$ is the trace of $\bA$, $\| \bA\|_{\text F}$ is the Frobenius norm of $\bA$, and $\bR$ is the correlation matrix of the SNPs (LD matrix). 

To check suspect pedigrees for hidden relatedness, one can compare theoretical kinships $\phi_{ij}$ and empiric kinships $\hat\phi_{ij}$. It is convenient to put these on a common scale by subjecting them to an approximate variance-stabilizing transformation. RA Fisher considered the simpler problem of comparing an ordinary covariance matrix $\bSigma = (\sigma_{ij})$ to a sample covariance matrix $\bS=(s_{ij})$. Under an assumption of normality, he argued \cite{fisher1915frequency,fisher1921probable} that the quantity
\begin{eqnarray*}
\tanh^{-1}\left(\frac{s_{ij}}{\sqrt{s_{ii}s_{jj}}}\right)
-\tanh^{-1}\left( \frac{\sigma_{ij}}{\sqrt{\sigma_{ii}\sigma_{jj}}}\right)
\end{eqnarray*} 
is approximately normal with mean $0$ and variance $(K-3)^{-1}$, where $K$ is the sample size, and $\tanh^{-1}$ is the inverse hyperbolic tangent function. By analogy, we subject the GRM matrix or one of its variants to Fisher's transformation and order the discrepancies from least to greatest in absolute value. The \openmendel\ MendelKinship.jl tutorial explains in a concrete example how transformation identifies outlier pairs.

\section{Variance Component Models}
\label{Variance Component Models}

Association studies are subject to the effects of unmeasured confounding. The most common confounder is ethnic ancestry \cite{astle2009population,helgason2005icelandic,knowler1988gm3}, which arises when  both trait values and marker allele frequencies  differ by region of origin.  Ancestry informative markers are particularly prone to show up as false positives in a na\"ive GWAS \cite{rosenberg2003informativeness}.   Currently there are two general adjustments for ethnic ancestry. The first approach uses either a few principal components of the GRM matrix \cite{patterson2006population,Price06EigenStrat,zhu2002association} or estimated ancestry proportions \cite{alexander2009admixture,pritchard2000inference} as fixed effects. The second approach explicitly accounts for the correlation between subjects by including an estimate of the kinship matrix, e.g. the GRM matrix, as a random effect in a variance components model.  When reliable pedigrees are available, the second approach is analogous to positing the theoretical kinship matrix as a random effect  \cite{boerwinkle1987use}. Because the theoretical kinship matrix does not capture hidden correlations, inclusion of the one of the SNP based estimates of the kinship matrix is usually preferred. 

In any event, the variance components model $\by \sim N(\bX\bbeta, \sum_{j=1}^k \sigma_j^2 \bV_j)$ figures prominently in genomewide association testing \cite{falconer1996c,lange2003mathematical}. 
In this model $\bbeta$ are the fixed effects of covariates $\bX$ and $\sigma_j^2$ is the variance of the jth random effect.  Estimation of the parameter vectors $\bbeta$ and $\bsigma^2 =(\sigma_1^2,\ldots,\sigma_k^2)^t$ has been the subject of intense study for decades. Most statisticians opt for maximum likelihood or restricted maximum likelihood. In the linear mixed model, the covariance matrices $\bV_j$ factor as $\bU_j\bU_j^t$. The factored form is advantageous if $\bU_j$ is $n \times r_j$ with $r_j$ small. In the absence of low rank structure, one can take $\bU_j$ to be the Cholesky factor of $\bV_j$.  

The covariance model $\bW= 2 \sigma^2_a\bPhi + \sigma^2_e \bI$ corresponds to polygenic background $(\sigma^2_a)$ plus random noise $(\sigma^2_e)$. The kinship matrix here can be theoretical or empirical. The model is overly simplistic but widely applied due to its computational tractability. It omits dominance effects, shared environment, and parent of origin effects, among other things. Calculation of the inverse and determinant of $\bW$ is the rate limiting step in estimation. In the simple polygenic model, a good tactic is to first calculate the spectral decomposition $\bO\bD\bO^t$ of $\bPhi$, where $\bD$ is a diagonal matrix. One can then exploit the formulas $\det \bW = \det (\sigma_a^2\bD + \sigma_e^2 \bI)$ and $\bW^{-1}  = \bO (\sigma_a^2\bD + \sigma_e^2 \bI)^{-1}\bO^t$. The indicated determinant and inverse of the diagonal matrix are trivial to compute \cite{kang2010variance,lippert2011fast,svishcheva2012rapid}.

Our program VarianceComponentModels.jl incorporates this spectral decomposition tactic. It also treats more realistic models with multiple variance components and multivariate traits. For estimation we have compared Fisher scoring and the EM algorithms long familiar to computational statisticians. We have also explored a new MM algorithm that alternates updates of $\bbeta$ and $\bsigma^2$ \cite{ZhouHuZhouLange2015VCMM}. The normal equation update of $\bbeta$ in our algorithm is
\begin{eqnarray*}
\bbeta_{n+1} & = & (\bX^t\bW_n^{-1}\bX)^{-1}\bX^t\bW_n^{-1}\by,
\end{eqnarray*}
where $\bW_n$ is the value of $\sum_{j=1}^k \sigma_j^2 \bV_j$ at the current estimate of $\bsigma^2$. The variance component updates are
\begin{eqnarray}
\sigma_{n+1,i}^{2} & = & \sigma_{ni}^{2} \sqrt{\frac{(\by - \bX\bbeta_{n+1})^t\bW_n^{-1} 
\bV_i \bW_n^{-1} (\by - \bX \bbeta_{n+1})}{\tr (\bW_n^{-1}  \bV_i)}}.
\label{variance_component_update}
\end{eqnarray}
The MM algorithm converges faster than the standard EM algorithm.  
Fisher scoring requires fewer iterations to converge but substantially 
more effort per iteration, particularly in high dimensions. Both the EM and MM algorithms can be accelerated by quasi-Newton extrapolation
\cite{zhou2011quasi}.

The VarianceComponentModels.jl module serves as a convenient vehicle for other genetic applications. One example is Mendelian randomization (MR).  Observational studies often find an association between a biomarker or expression (or methylation) level at a particular locus and a quantitative trait.  The goal of MR is to assess the statistical support for this ``exposure" as a cause of the trait, as opposed to reverse causality or confounding \cite{burgess2015mendelian}. Our Mendelian randomization tutorial for continuous traits demonstrates the value of modularized genetic software such as VarianceComponentModels.jl. 

When there are many loci to test, we \cite{Clark2016MFGScoreTest,Zhou2017FastScore} and others \cite{amin2007genomic,chen2007family,kang2010variance,lippert2011fast} have employed score tests or their equivalents in variance component models. Score tests are much faster than likelihood ratio tests (LRTs) because score tests require the likelihood to be maximized only under the null hypothesis. In contrast, LRTs require the likelihood be maximized both under the null and alternative hypotheses.  When the null hypothesis is the same for all loci tested, this can amount to substantial savings. These score tests are easily extended to include maternal genetic effects and maternal-offspring genetic interaction as fixed effects \cite{Clark2016MFGScoreTest}.  Although most software programs implementing the score test adopt the simple covariance model $2 \sigma^2_a\bPhi + \sigma^2_e \bI$ , in principle other variance components such as household effects can be included.

Our recent analysis of the GWAS data from the COPDGene study (\url{http://www.copdgene.org}) exemplifies the vast performance gain and yet ease of use of a typical \openmendel\ workflow in genetic heritability analysis of a realistically large data set \cite{Zhou18MMVCCOPD}. The data are available from NIH dbGap under phs000179.v5.p2. The steps are: (1) load the binary genotypes of 6,670 individuals at 630,860 SNPs, (2) compute summary statistics on the SNPs, (3) impute missing genotypes, (4) calculate the empirical kinship matrix, (5) load 13 phenotypes, (6) estimate the heritability of each phenotype, (7) estimate the coheritability of each pair of phenotypes, and (8) fit a joint model to all 13 phenotypes. All these steps are performed in a single interactive Julia environment on a common laptop computer. Typically such an analysis pipeline would require running at least five separate programs on a Linux machine. 

\begin{figure}[ht]
\begin{center}
$$
\begin{array}{cc}
\includegraphics[width=3.5in]{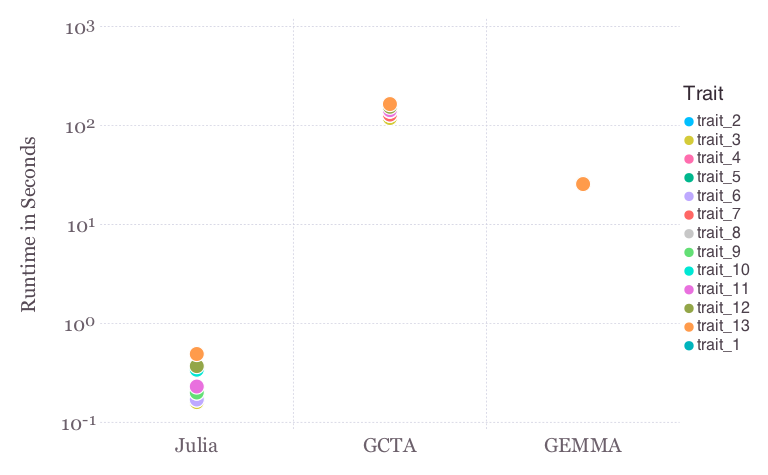} \\
\\
\includegraphics[width=3.5in]{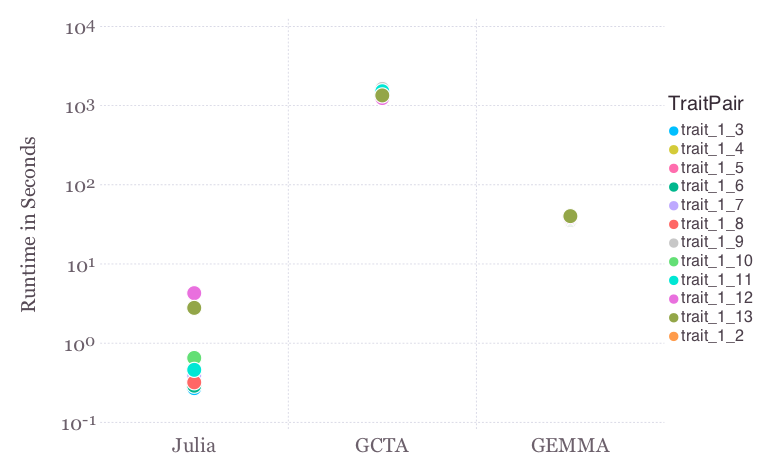}
\end{array}
$$
\end{center}
\caption{Comparison of  the \openmendel\  VarianceComponentModels.jl implementation with GCTA (C++) and GEMMA (C++) for fitting a univariate variance component model
$\Ybf \sim N(\zerobf, \sigma_a^2 \bS + \sigma_e^2 \Ibf)$ (top panel) and a bivariate variance component model
$ \vec{\Ybf}\sim N(\zerobf, \bSigma_a \otimes \bS + \bSigma_e \otimes \Ibf)$ (bottom panel).
GEMMA and \openmendel\ runtimes exclude the eigen-decomposition of $\bS$, which is pre-computed.}
\label{fig:julia-vs-gcta-gemma}
\end{figure}

In our experience, a pure Julia\ computation is often faster than the corresponding computation in a low-level language such as C, and much faster than any other high-level language such as R or Python. Figure \ref{fig:julia-vs-gcta-gemma} compares the speed of fitting large-scale variance component models in our  VarianceComponentModels.jl module to the two cutting edge programs GCTA \cite{yang2011gcta} and GEMMA \cite{ZhouStephens14GEMMA}, both implemented in C++. In this example, there are two variance components, one for additive genetic effects and one for environmental effects. To make a fair comparison, the genetic relationship matrix $\bS$ was pre-computed using the GCTA software. There are 13 continuous phenotypes. For both univariate (top panel) and bi-variate (bottom panel) models, we observe between 5 and 100 fold speedup over GEMMA and even more over GCTA. In all cases, the final log-likelihoods by \julia\ match those by GCTA and GEMMA to the third digit. 

The current versions of GCTA and GEMMA are only available for the x86 64-bit Linux operating system, while \julia, and thus \openmendel, are available on all common systems. It is remarkable that a cross-platform, interactive, high-level language such as \julia\ can achieve such excellent computational efficiency.

\section{SNP-Set Analysis}
\label{sec:snp-set}

SNP-set analysis, or pathway-based analysis \cite{wang2007pathway}, is a powerful, widely-used strategy in sequencing studies. SNPs are grouped into sets to be examined for association with a certain phenotype. This analysis has been shown to have increased power over individual SNP analysis, especially for identifying rare variant associations \cite{Lee14RVSurvey}. 

Two types of SNP-set analyses are under active development within \openmendel. The VarianceComponentTests.jl module implements different approaches for testing a set of markers as random effects. Notably the sequence kernel association test (SKAT) \cite{wu2011rare} is the first method to incorporate the generalized linear mixed model in testing the effect of a set of variants on a quantitative or dichotomous trait. Our recent work on exact tests \cite{zhou2016boosting} boosts the power of SKAT on small samples. 

In contrast to marginal SNP-set analysis, an alternative approach is subset selection in a joint model $\by \sim N(\bX\bbeta, \sum_{j=1}^m \sigma_j^2 \bV_j + \sigma_0^2 \bI)$, where $\bV_j$ is the kernel matrix for the $j$th SNP-set and the $\sigma_j^2$ for $j \ge 1$ are the variance components subject to selection. Variance component selection is achieved by minimizing the penalized log-likelihood 
\begin{equation*}
\begin{aligned}
-L(\boldsymbol{\beta}, \boldsymbol{\sigma^2})+\sum_{j=1}^m P_{\lambda}(\sigma_j),
\end{aligned}
\end{equation*}
where $L(\boldsymbol{\beta}, \boldsymbol{\sigma^2})$ is the log-likelihood function and $P_\lambda(\sigma_j)$ is a penalty function. Several penalties, including the ridge, the lasso, the smoothly clipped absolute deviation (SCAD), and the minimax concave penalty (MCP), are implemented. The MM update \eqref{variance_component_update} generalizes to penalized estimation because the variance components $\sigma_j^2$ are nicely separated in the surrogate function \cite{ZhouHuZhouLange2015VCMM}.

\section{Simulation Utilities}
\label{sec:simulation}

Simulation is vital in demonstrating the accuracy and power of  new statistical methods. It is also important 
in designing genetic studies, where overly simplistic assumptions can lead to low power.  Although there are a number of simulators already available \cite{liu2008survey,schaffer2011coordinated,yuan2012overview}, there is plenty of room for improvement. The unified nature of the  \openmendel\ environment makes it easy to craft code for simulating traits conditional on genotypes under any generalized linear model (GLM) or generalized linear mixed models (GLMM). 

At the time that this article was written, the MendelTraitSimulate.jl option was under development. In its current version, we accommodate study designs involving both unrelateds and multigenerational families. We allow the user to specify both fixed and random effects for simulated univariate or multivariate traits. The simulated traits can be based on arbitrary functions of the provided covariates. By default, the program will use the PLINK format and make appropriate calls to  SnpArrays.jl and VCFTools.jl.

\section{Tutorials}
\label{sec:tutorials}


Accompanying this article we have prepared a collection of tutorials via Jupyter Notebooks to demonstrate interactive genetic analysis using \openmendel\ packages (\url{https://github.com/OpenMendel/Tutorials}). These include (1) PLINK binary data input, summary statistics, filtering, and visualization, (2) kinship calculation and comparison, (3) population GWAS, (4) iterative hard thresholding for GWAS, (5) heritability estimation, (6) Mendelian randomization, (7) GWAS based on linear mixed models, (8) SNP-set analysis, and soon to come (9) trait simulation. These tutorials will adapt to and grow with the expanding \openmendel\ ecosystem.

\section{Discussion}

Readers may be familiar with our existing statistical package \mendel\ \cite{lange2013mendel}. Although \mendel\ possesses many advantages, our goal going forward is not to modernize it, but to create an entirely new open source platform. Although \mendel\ is free, it is not open source. The Fortran language underlying it is also antiquated. Fortran lacks the supporting libraries of R and Matlab, its graphics functionality is nil, 
it neglects crucial statistical and linear algebra tools, and its code is needlessly verbose. 

For the sake of brevity, we have not discussed many \openmendel\ modules. Omitted modules include: (1) discovery of ancestry informative markers, (2) estimation of allele frequencies from pedigree data, (3) testing for transmission disequilibrium by the gamete competition model, (4) random genotype generation by gene dropping, (5) genetic counseling, (6) two-point linkage analysis, (7) location scores for linkage analysis, and (8) function optimization by recursive quadratic programming.  Table \ref{table:openmendel-packages} lists the currently available \openmendel\ analysis options and utility packages as well as those soon to be released as part of the \openmendel\ project.   

\begin{table}[htb]
\begin{center}
{\small
\begin{tabular}{rl}
\toprule
\openmendel\ Option  & Description\\
\midrule
{\tt MendelAimSelection.jl} & Selects the most informative SNPs for predicting ancestry \\
{\tt MendelEstimateFrequencies.jl} & Estimates allele frequencies from pedigree data\\
{\tt MendelGameteCompetition.jl} & Tests for association under the gamete competition model \\
{\tt MendelGeneticCounseling.jl} & Computes risks in genetic counseling problems \\
{\tt MendelGWAS.jl} & Tests for association in genome-wide data \\
{\tt MendelIHT.jl} & GWAS using Iterative Hard Thresholding (forthcoming)\\
{\tt MendelImpute.jl} & Genotype imputation (forthcoming)\\
{\tt MendelKinship.jl} & Computes kinship and other identity coefficients \\
{\tt MendelLocationScores.jl} & Maps a trait via the method of location scores \\
{\tt OrdinalGWAS.jl} & Implements GWAS for ordinal categorial phenotypes \\
{\tt MendelTwoPointLinkage.jl} & Implements two-point linkage analysis \\
\midrule
{\tt MendelBase.jl} & Base functions for \openmendel\ \\
{\tt MendelGeneDropping.jl} & Simulates genotypes based on pedigrees \\
{\tt MendelSearch.jl} & Optimization routines \\
{\tt MendelTraitSimulate.jl} & Trait simulation using GLM and GLMM (forthcoming)\\
{\tt SnpArrays.jl} & Utilities for handling compressed storage of biallelic SNP data \\
{\tt VCFTools.jl} & Utilities for handling compressed storage of sequence data \\
{\tt VarianceComponentModels.jl} & Utilities for fitting and testing variance components models \\
\bottomrule
\end{tabular}
}
\caption{\julia\ packages in the \openmendel\ project.}
\label{table:openmendel-packages}
\end{center}
\end{table}

\openmendel\ is inspired by a vision of genomic analysis that extracts the maximum benefit from the world-wide increase in genetic data and exploits the promise of collaborative, parallel, and distributed computing. We are not alone in this vision.  As examples, notable strides have been made by HAIL (\url{https://hail.is/}) and TOPMed (\url{https://www.nhlbiwgs.org/awards}) in enabling large scale sequence analysis and the Ark data management system for health and biomedical research \cite{Bickerstaffe17Ark,Ranaweera18PedDataTools}. In our opinion, however, the barriers need to be lowered further to encourage more statistical geneticists and genetic epidemiologists to take part. The \julia\ language provides the ideal vehicles for this purpose. It is our hope that the \openmendel\ project will spark a global effort to build a computing platform equal to the challenges of $21$st-century genetic research.

\section*{Acknowledgements}

This work was funded by NIH Grants R01-GM53275, R01-HG006139, R01-GM105785, R01-HL135156 and T32-HG002536; NSF grant DMS-1052210; the UCSF Bakar Computational Health Sciences Institute; the UC Berkeley Institute for Data Sciences as part of the Moore-Sloan Data Sciences Environment Initiative; and the 2018 Google Summer of Code.

\bibliographystyle{plain}
\bibliography{review4}

\end{document}